# The electrocaloric response in Lanthanum-modified lead zirconate titanate ceramic

B. Asbani[1], J.-L. Dellis[1], A. Lahmar[1], M. Amjoud[2], Y. Gagou[1], D. Mezzane[2], Z. Kutnjak[3], R. Pirc[3], M. El Marssi[1], B. Rožič[3]

[1]*LPMC, Université de Picardie Jules Verne, 33 rue Saint-Leu, 80039 Amiens Cédex, France*
[2]*LMCN, F.S.T.G. Université Cadi Ayyad, BP 549, Marrakech, Morocco*
[3]*Jozef Stefan Institute, Jamova cesta 39, 1000 Ljubljana, Slovenia*

## Abstract

Over the past decades, there has been significant interest in new cooling technology based on the electrocaloric effect. The large electrocaloric effect observed in polymeric and inorganic ferroelectric materials made possible development of dielectric cooling devices of a new generation. We report a significant impact of annealing on the electrocaloric effect observed in 12/65/35 PLZT bulk ceramics. The electrocaloric data were obtained by direct measurements. Electrocaloric results confirm the existence of the significant electrocaloric response in this relaxor ferroelectric PLZT composition exceeding previously obtained electrocaloric values in perovskite relaxor ferroelectrics such as PMN-PT and x/65/35 PLZT ceramics.

**Keywords**: Ferroelectric, electrocaloric, lanthanum, dielectric, responsivity.

Corresponding author: brigita.rozic@ijs.si

## Introduction

The mechanism underlying the electrocaloric effect (ECE) is related to the change of thermal entropy that is compensating the change of dipolar entropy being stimulated by the electric field under adiabatic conditions [1-7]. Over the past decades there has been a particular interest in new cooling technology based on an electrocaloric effect, yet due to relatively small ECE observed previously in ferroelectrics no much progress was made [8-10]. A new stimulus was given to ECE research by the recent discovery of a giant electrocaloric effect in organic and inorganic ferroelectric materials [1-23]. Specifically, indirect ECE data indicated the giant electrocaloric effect in a thin $PbZr_{0.95}Ti_{0.05}O_3$ (PZT) [2] film and thick poly(vinylidene fluoride-trifluoroethylene) (P(VDF-TrFE)) copolymer films [3]. Numerous indirect and direct measurements performed on bulk samples demonstrated significant electrocaloric effect in various bulk and thin-film perovskite ferroelectric materials [1,12-24], i.e., in materials that have also been very interesting for application due to their large dielectric and electromechanical properties [25]. It was shown experimentally and theoretically that both positive and negative ECE exist in ferroelectric (FE) and antiferroelectric (AFE) materials [26-29]. Theoretical calculations and experiments show that enhanced electrocaloric effect can be found in perovskite relaxor ferroelectrics in the vicinity of the critical point [24]. Such diffuse critical point was found to exist also in relaxor compositions of La-modified lead zirconate titanate (x/65/35 PLZT) ceramics with 0.05<x<0.12 [24,30]. Such enhancement is related to the relatively large ECE observed in lead magnesium niobate lead titanate (PMN-10PT) with an electrocaloric temperature change of $\Delta T_{EC} = 3.45$ K [28]. Relatively large ECE of 2.25 K was also observed in bulk 8/65/35 PLZT ceramics [24] and ECE of 40 K in 8/65/35 PLZT thin films, however, at significantly higher electric fields [14]. Such large electrocaloric values observed in these two canonical relaxor ferroelectrics can already be exploited in electrocaloric cooling devices [23,32]. Recently, large ECE of 6 K was also reported in antiferroelectric composition (2/95/5) of PLZT thin film ceramics [33-35].

In this work, we investigate by direct electrocaloric measurements the temperature and electric field dependence of the electrocaloric response in $Pb_{1-x}La_x(Zr_yTi_{1-y})_{1-x/4}O_3$ bulk ceramics with $x$=0.12 and $y$=0.65 (denoted by 12/65/35 PLZT ceramics). The electrocaloric effect was investigated in the temperature range near the room temperature, i.e., from 300 K up to 440 K. The results demonstrated importance of the accumulated stress annealing during the sample preparation well above the paraelectric to ferroelectric phase transition for electrocaloric response enhancement.

Calculations performed in Refs. [5,6,24] by using the self-consistent equation

$$\frac{\Delta T_{EC}}{T} = exp\left\{\frac{P^2(0,T) - P^2(E, T + \Delta T)}{2C_{ph}(T)}\right\}$$

demonstrated the electrocaloric temperature change $\Delta T_{EC}$ enhancement in the vicinity of the critical point for simple ferroelectrics or relaxor ferroelectrics. Here, the polarization is obtained from the equations of state based on the free energy expressed as [5]

$$f = f_0 + \frac{1}{2}aP^2 + \frac{1}{4}bP^4 + \frac{1}{6}cP^6 - PE,$$

where the Landau type expansion coefficients are denoted by $a = a_1(T - T_0)$, $b$ and $c$, with $T_0$ the paraelectric-to-ferroelectric transition temperature in zero field. The detailed calculations for relaxor ferroelectrics and ferroelectrics have demonstrated that far from the critical point, the maximum of the ECE response is expected at the ferroelectric phase transition [5,6,18]. In some relaxor ferroelectrics such as PMN and x/65/35 PLZT ceramics with x above 0.07, the ferroelectric phase is induced only by the strong enough electric field [25,36-38]. Here, we will investigate the 12/65/35 PLZT ceramics with the relaxor ferroelectric properties that exhibit smeared dielectric peak at about 320 K [39]. Computational details of the above calculations are available in [40].

**Experiment**

The ECE was determined by direct electrocaloric measurements in bulk 12/65/35 PLZT ceramics that were prepared by a conventional mixed-oxide method followed by hot-press sintering at 1250 °C and final half-an-hour annealing at 600 °C [35,37]. All prepared samples were single-phase perovskites determined by the X-ray diffraction analysis [37]. This material is known to possess interesting dielectric and pyroelectric properties [37].

The resulting PLZT platelets were cut and thinned down to the thickness of 140 µm. After cutting and thinning this sample was measured as a virgin sample, i.e., without additional annealing. Another 140 µm thick platelet was additionally thinned to 68 µm thickness and was once more annealed for one hour at 600 °C. Gold electrodes covered both surfaces of all samples. The details of the direct electrocaloric measurement method and data analysis are described in Refs. [11,12,24]. Specifically, the direct electrocaloric measurement method was based on modified high-resolution calorimeter that allows precise temperature stabilization of the bath (within ± 0.1 mK) and high-resolution measurements of the sample temperature variations due to the electrocaloric effect induced by the change of the applied electric field [24]. A small bead thermistor was used as a temperature sensor.

**Results and discussion**

Figure 1 shows the real part of the complex dielectric constant obtained at 10 kHz as a function of temperature for both 140 µm virgin (a) and 68 µm (b) annealed PLZT ceramics. Significant hysteresis was observed in virgin 140 µm thick sample, and a much smaller value of the dielectric constant than that observed in annealed 68 µm thick sample. The magnitude of the dielectric constant of the annealed sample is surprisingly higher than previously reported for 12/65/35 PLZT composition [37].

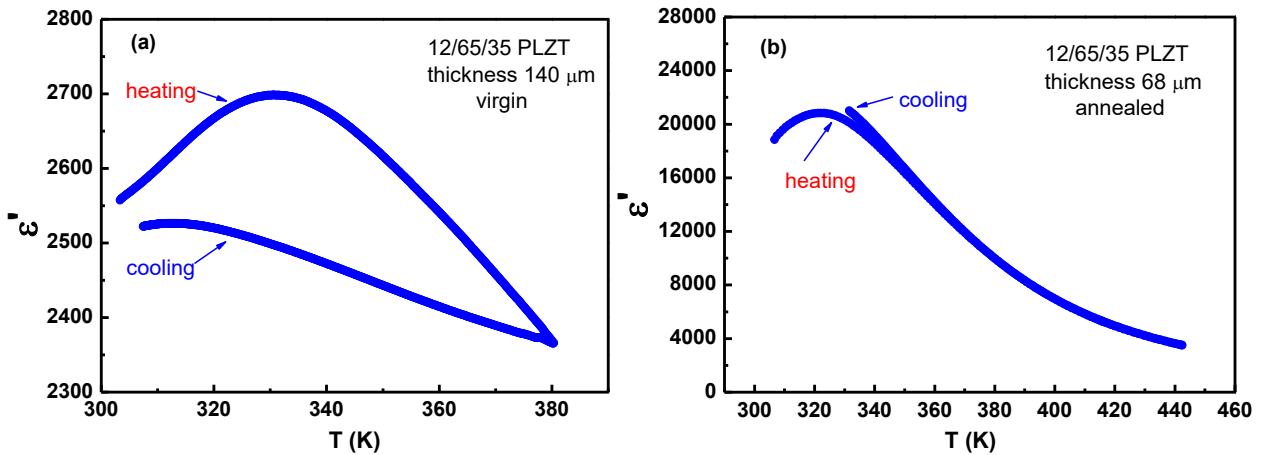

**Fig. 1.** Temperature dependence of the real part of the dielectric constant ε' measured at 10 kHz in bulk virgin 140 µm thick 12/65/35 PLZT ceramics (a) and 68 µm annealed bulk 12/65/35 PLZT ceramics (b).

The dielectric temperature profiles in both samples clearly show the relaxor ferroelectric character with the broad diffuse dielectric peak near 320 K, typical for the 12/65/35 PLZT ceramics. Visible hysteresis between cooling and heating runs indicate history effects in virgin sample probably related to partial annealing of the imprinted stress field. As shown in Ref. [24] the induced critical point by the electric field can provide enhancement of the ECE response. The temperature profiles of the ECE temperature change $\Delta T_{EC}$ for both virgin and annealed PLZT samples are shown in Figure 2 at a few selected amplitudes of the electric field. In order to preserve the sample and not to exceed the breakdown field, the electric-field amplitude was limited to a maximum of 90 kV/cm.

The temperature profiles of the ECE specific entropy change $\Delta s$ for both virgin and annealed PLZT samples are shown in Figure 3. Specific entropy change $\Delta s$ is calculated by scaling the total measured ECE entropy change by the mass of the sample.

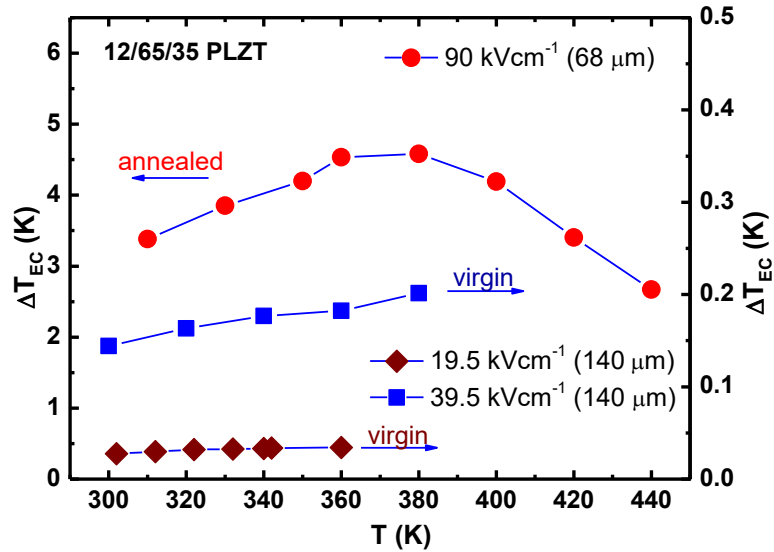

**Fig. 2.** The ECE temperature change $\Delta T_{EC}$ in virgin and annealed 12/65/35 PLZT bulk ceramics as a function of temperature for several amplitudes of the electric field.

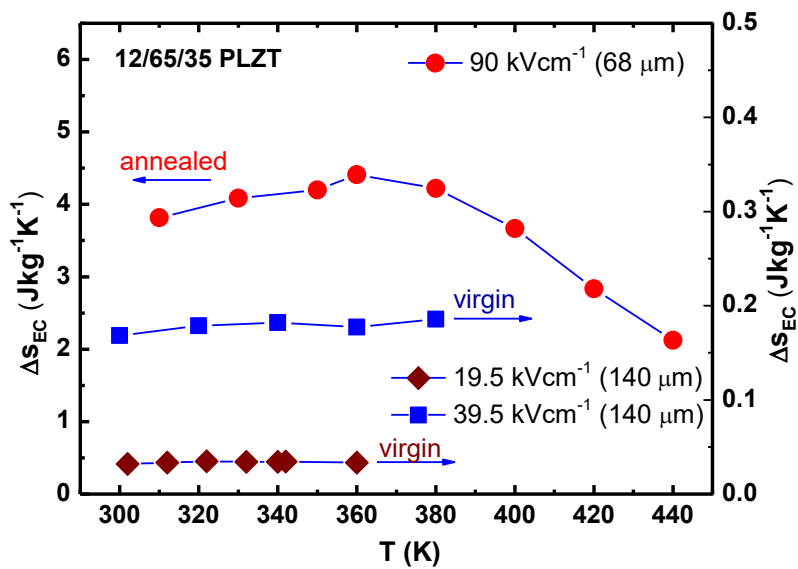

**Fig. 3.** The ECE specific entropy change Δ*s* in virgin and annealed 12/65/35 PLZT bulk ceramics as a function of temperature for several amplitudes of the electric field.

Results presented in Figs. 2 and 3 demonstrate that the ECE is positive in a whole measured temperature range, even for lower electric fields as expected for the electric field induced ferroelectric phase in relaxor ferroelectrics [5,6,28,29].

**Discussion**

The temperature ECE profiles of the ECE response determined in 12/65/35 PLZT bulk ceramics are qualitatively similar to those observed in other La-modified relaxor ferroelectric PLZT compounds, i.e., temperature profiles with a smeared maximum at the electric-field induced FE phase transition temperature. The maximum value of $\Delta T_{EC}$ observed in virgin 12/65/35 PLZT bulk ceramics was more than an order of magnitude smaller (0.2 K at 360 K and 39.5 kV/cm) than in annealed 12/65/35 PLZT bulk ceramics (4.58 K at 360 K and 90 kV/cm). It appears that in virgin sample imprinted stress field precludes the existence of the critical point, thus suppressing the enhancement of the ECE response. In contrast, in the annealed 12/65/35 PLZT ceramics large ECE values demonstrate significant enhancement of the ECE response at the FE transition. It should be noted that by increasing the applied electric field, the FE transition is shifted toward higher temperatures as observed in other perovskite ferroelectrics [24]. The above maximum value of $\Delta T_{EC}$ measured in annealed 12/65/35 PLZT ceramics exceeds previously published values in all La-modified x/65/35 PLZT compositions [1,11,24]. It is even exceeding the value of 3.45 K observed in PMN-10PT but at the much higher field of 140 kV/cm [31]. We argue that such ECE enhancement for relatively modest electric fields is most likely due to the critical point induced by the electric field. The value of the electrocaloric responsivity *ΔT/ΔE*=5.09x10$^{-7}$ Km/V measured in annealed 12/65/35 PLZT ceramics at 360 K, is one of the largest values observed in perovskite ferroelectrics.

**Conclusion**

In conclusion, we have explored by direct electrocaloric measurements the temperature and electric field dependence of the electrocaloric response in bulk 12/65/35 PLZT ceramics. The impact of the annealing on the magnitude of the electrocaloric effect was investigated in the temperature range near the room temperature, i.e., from 300 K up to 440 K. Relatively small value of ECE response not exceeding 0.2 K at 39.5 kV/cm were found in the virgin 12/65/35 PLZT sample. In contrast, in the annealed 12/65/35 PLZT sample one of the highest reported values in bulk perovskites of $\Delta T_{EC}$ = 4.58 K at 360 K and 90 kV/cm was observed due to the critical point enhancement, which seems to be precluded by the imprinted stress field in the virgin sample. The results demonstrated the importance of the proximity of the critical point for electrocaloric response enhancement.

**Acknowledgments**

The authors thank the project 778072 — ENGIMA — H2020-MSCA-RISE-2017, Slovenian Research Agency grant J1-9147 and program P1-0125. Authors would also like to acknowledge critical help in sample preparation by Jena Cilenšek and Silvo Drnovšek.

# References

[1] M. Valant, Electrocaloric materials for future solid-state refrigeration technologies, Prog. Mater. Sci. 57 (2012) 980–1009. doi: 10.1016/j.pmatsci.2012.02.001

[2] A. S. Mischenko, Q. Zhang, J. F. Scott, R. W. Whatmore, N. D. Mathur, Giant Electrocaloric Effect in Thin-Film PbZr0.95Ti0.05O3, Science 311 (2006) 1270–1271. doi:10.1126/science.1123811

[3] B. Neese, B. Chu, S.-G. Lu, Y. Wang, E. Furman, and Q.M. Zhang, Large Electrocaloric Effect in Ferroelectric Polymers Near Room Temperature, Science 321 (2008) 821–823. doi:10.1126/science.1159655

[4] J. F. Scott, Applications of Modern Ferroelectrics, Science 315 (2007) 954–959. doi: 10.1126/science.1129564

[5] R. Pirc, Z. Kutnjak, R. Blinc, Q. M. Zhang, Electrocaloric effect in relaxor ferroelectrics, J. Appl. Phys. 110 (2011) 074113. https://doi.org/10.1063/1.3650906

[6] R. Pirc, Z. Kutnjak, R. Blinc, Q. M. Zhang, Upper bounds on the electrocaloric effect in polar solids, Appl. Phys. Lett. 98 (2011) 021909. https://doi.org/10.1063/1.3543628

[7] I. Ponomareva and S. Lisenkov, Bridging the Macroscopic and Atomistic Descriptions of the Electrocaloric Effect, Phys. Rev. Lett. 108 (2012), 167604. DOI: https://doi.org/10.1103/PhysRevLett.108.167604

[8] A. S. Mischenko, Q. Zhang, J. F. Scott, R. W. Whatmore, N. D. Mathur, Giant electrocaloric effect in the thin film relaxor ferroelectric 0.9PbMg1∕3Nb2∕3O3–0.1PbTiO30.9PbMg1∕3Nb2∕3O3–0.1PbTiO3 near room temperature, Appl. Phys. Lett. 89 (2006), 24291. https://doi.org/10.1063/1.2405889

[9] G. Akcay, S. P. Alpay, G. A. Rosseti, J. F. Scott, Influence of mechanical boundary conditions on the electrocaloric properties of ferroelectric thin films, J. Appl. Phys. 103 (2008), 024104. https://doi.org/10.1063/1.2831222

[10] D. Guyomar, G. Sebald, B. Guiffard, L. Seveyrat, Ferroelectric electrocaloric conversion in 0.75(PbMg1/3Nb2/3O3)–0.25(PbTiO3) ceramics, J. Phys. D Appl. Phys. 39 (2006), 4491– 4496. https://doi.org/10.1088/0022-3727/39/20/02

[11] G. Sebald, S. Pruvost, L. Seveyrat, L. Lebrun, D. Guyomar, B. Guiffard, Electrocaloric properties of high dielectric constant ferroelectric ceramics, J. Europ. Ceram. Soc. 27 (2007), 4021. https://doi.org/10.1016/j.jeurceramsoc.2007.02.088

[12] N. D. Mathur, Future Trends in Electrocalorics Materials, in: T. Correia, Q. Zhang (Eds.), Electrocaloric Mater., Springer Berlin Heidelberg, Berlin, Heidelberg, 2014: pp. 251–253. https://link.springer.com/chapter/10.1007/978-3-642-40264-7_10 (accessed May 24, 2019).

[13] Z. Kutnjak, B. Rožič and R. Pirc, Electrocaloric Effect: Theory, Measurements, and Applications, Wiley Encyclopedia of Electrical and Electronics Engineering (Wiley) 2015: pp. 1–19. https://doi.org/10.1002/047134608X.W8244

[14] S.G. Lu, B. Rožič, Q.M. Zhang, Z. Kutnjak, X. Li, E. Furman, L.J. Gorny, M. Lin, B. Malič, M. Kosec, R. Blinc, and R. Pirc, Organic and inorganic relaxor ferroelectrics with giant electrocaloric effect, Appl. Phys. Lett. 97 (2010), 162904. https://doi.org/10.1063/1.3501975

[15] X. Moya, S. Kar-Narayan, N.D. Mathur, Caloric materials near ferroic phase transitions, Nat. Mater. 13 (2014) 439–450. doi:10.1038/nmat3951.

[16] J. Peräntie, J. Hagberg, A. Uusimäki, H. Jantunen, Electric-field-induced dielectric and temperature changes in a ⟨011⟩-oriented Pb(Mg1/3Nb2/3)O3-PbTiO3 single crystal, Phys. Rev. B 82 (2010), 134119. https://doi.org/10.1103/PhysRevB.82.134119

[17] Y. Bai, X. Han, and L. Qiao, Optimized electrocaloric refrigeration capacity in lead-free (1−x)BaZr0.2Ti0.8O3-xBa0.7Ca0.3TiO3 ceramics, Appl. Phys. Lett. 102 (2013), 252904. https://doi.org/10.1063/1.4810916

[18] N. Novak, Z. Kutnjak, and R. Pirc, High-resolution electrocaloric and heat capacity measurements in barium titanate, EPL 103 (2013), 47001. https://doi.org/10.1209/0295-5075/103/47001

[19] B. Asbani, J.-L. Dellis, A. Lahmar, M. Courty, M. Amjoud, Y. Gagou, K. Djellab, D. Mezzane, Z. Kutnjak, M. El Marssi, Lead-free Ba$_{0.8}$Ca$_{0.2}$(Zr$_x$Ti$_{1-x}$)O$_3$ ceramics with large electrocaloric effect, Appl. Phys. Lett. 106 (2015) 042902. doi:10.1063/1.4906864.

[20] Y. Bai, G. Zheng, S. Shi, Direct measurement of giant electrocaloric effect in BaTiO3BaTiO3 multilayer thick film structure beyond theoretical prediction, Appl. Phys. Lett. 96 (20109, 192902. https://doi.org/10.1063/1.3430045

[21] H. Kaddoussi, A. Lahmar, Y. Gagou, B. Asbani, J.-L. Dellis, G. Cordoyiannis, B. Allouche, H. Khemakhem, Z. Kutnjak, M. El Marssi, Indirect and direct electrocaloric measurements of $(Ba_{1-x}Ca_x)(Zr_{0.1}Ti_{0.9})O_3$ ceramics (x = 0.05, x = 0.20), J. Alloy. Comp. 667 (2016), 198–203. https://doi.org/10.1016/j.jallcom.2016.01.159

[22] B. Asbani, J.-L. Dellis, Y. Gagou, H. Kaddoussi, A. Lahmar, M. Amjoud, D. Mezzane, Z. Kutnjak, M. El Marssi, Electrocaloric effect in Ba0.2Ca0.8Ti0.95Ge0.05O3 determined by a new pyroelectric method, EPL 111 (2015), 57008. https://doi.org/10.1209/0295-5075/111/57008

[23] X. Moya, E. Defay, V. Heine, N. D. Mathur, Too cool to work, Nat. Phys. 11(2015), 202–205. Doi:10.1038%2Fnphys3271

[24] B. Rožič, M. Kosec, H. Uršič, J. Holc, B. Malič, Q. M. Zhang, R. Blinc, R. Pirc, Z. Kutnjak, Influence of the critical point on the electrocaloric response of relaxor ferroelectrics, J. Appl. Phys. 110 (2011), 064118. https://doi.org/10.1063/1.3641975

[25] Z. Kutnjak, J. Petzelt, R. Blinc, The giant electromechanical response in ferroelectric relaxors as a critical phenomenon, Nature 441(2006), 956-959. https://doi.org/10.1038/nature04854

[26] Y. Bai, G. Zheng, S. Shi, Abnormal electrocaloric effect of Na0.5Bi0.5TiO3–BaTiO3 lead-free ferroelectric ceramics above room temperature, Mater. Res. Bull. 46, 1866–1869 (2011). https://doi.org/10.1016/j.materresbull.2011.07.038

[27] B. Peng, H. Fan, Q. Zhang, A Giant Electrocaloric Effect in Nanoscale Antiferroelectric and Ferroelectric Phases Coexisting in a Relaxor Pb0.8Ba0.2ZrO3 Thin Film at Room Temperature, Adv. Funct. Mater. 23 (2013), 2987–2992. https://doi.org/10.1002/adfm.201202525

[28] R. Pirc, B. Rožič, J. Koruza, B. Malič, and Z. Kutnjak, Negative electrocaloric effect in antiferroelectric PbZrO3, Europhys. Lett. 107 (2014), 17002. https://doi.org/10.1209/0295-5075/107/17002

[29] R. Pirc, B. Rožič, J. Koruza, G. Cordoyiannis, B. Malič, Z. Kutnjak, Anomalous dielectric and thermal properties of Ba-doped PbZrO3 ceramics, J. Phys.: Condens. Matter 27 (2015), 455902. https://doi.org/10.1088/0953-8984/27/45/455902

[30] N. Novak, R. Pirc, Z. Kutnjak, Diffuse critical point in PLZT ceramics, EPL 102 (2013), 17003. https://iopscience.iop.org/article/10.1209/0295-5075/103/47001

[31] M. Vrabelj, H. Uršič, Z. Kutnjak, B. Rožič, S. Drnovšek, A. Benčan, V. Bobnar, L. Fulanović, B. Malič, Large electrocaloric effect in grain-size-engineered 0.9Pb(Mg1/3Nb2/3)O3–0.1PbTiO3, J. Eur. Ceram. Soc. 36 (2016), 75– 80. https://doi.org/10.1016/j.jeurceramsoc.2015.09.031

[32] U. Plaznik, A. Kitanovski, B. Rožič, B. Malič, H. Uršič, S. Drnošek, J. Cilenšek, M. Vrabelj, A. Poredoš, Z. Kutnjak, Bulk relaxor ferroelectric ceramics as a working body for an electrocaloric cooling device, Appl. Phys. Lett. 106 (2015), 043903. https://doi.org/10.1063/1.4907258

[33] Y. Mendez-González, A. Peláiz-Barranco, T. Yang, J. D. S. Guerra, Enhanced electrocaloric effect in La-based PZT antiferroelectric ceramics, Appl. Phys. Lett. 112 (2018), 122904. https://doi.org/10.1063/1.5018431

[34] G. Wenping, Y. Liu, X. Meng, L. Bellaiche, J. F. Scott, B. Dkhil, A. Jiang, Antiferroelectric Thin Films: Giant Negative Electrocaloric Effect in Antiferroelectric La-Doped Pb(ZrTi)O3 Thin Films Near Room Temperature, Advanced Materials 27 (2015), 3165. DOI: 10.1002/adma.201570137

[35] Z. Xu, Xunhu Dai, D. Viehland, Incommensuration in La-modified antiferroelectric lead zirconate titanate ceramics, Appl. Phys. Lett. 65 (1994), 3287. https://doi.org/10.1063/1.112441

[36] R. Farhi, M. El Marssi, A. Simon, J. Ravez, A Raman and dielectric study of ferroelectric Ba(Ti1-xZrx)O3 ceramics, Eur. Phys. J. B 9 (1999), 599–604. https://doi.org/10.1007/s100510050803

[37] Z. Kutnjak, C. Filipič, R. Pirc, A. Levstik, R. Farhi, M. El Marssi, Slow dynamics and ergodicity breaking in a lanthanum-modified lead zirconate titanate relaxor system, Phys. Rev. B 59 (1999), 294. https://doi.org/10.1103/PhysRevB.59.294

[38] V. Bobnar, Z. Kutnjak, R. Pirc, A. Levstik, Electric-field–temperature phase diagram of the relaxor ferroelectric lanthanum-modified lead zirconate titanate, Phys. Rev. B 60 (1999), 6420. DOI:https://doi.org/10.1103/PhysRevB.60.6420

[39] M. El Marssi, R. Farhi, J.-L. Dellis, M. D. Glinchuk, L. Seguin, D. Viehland, Ferroelectric and glassy states in La-modified lead zirconate titanate ceramics: A general picture, J. Appl. Phys. 83 (1998), 5371. https://doi.org/10.1063/1.367366
[40] Open access computational scripts are available online at https://github.com/lukyanc/Electrocaloric.